\documentclass[sigconf]{acmart}

\renewcommand\footnotetextcopyrightpermission[1]{} 
\AtBeginDocument{%
  }

\copyrightyear{2022}
\acmYear{2022}
\setcopyright{rightsretained}
\acmConference[KDDCup '22]{To be presented at KDD Cup 2022 Workshop:
ESCI Challenge for Improving Product Search}{August 17, 2022}{Washington, DC, USA}
\acmBooktitle{KDD Cup 2022 Workshop: ESCI Challenge for Improving Product Search, August 17, 2022, Washington, DC, USA}
\acmDOI{}
\acmISBN{}

\begin{document}

\title{Predicting Query-Item Relationship using Adversarial Training and Robust Modeling Techniques}

\author{Min Seok Kim}
\affiliation{%
  \institution{Kakao Corporation}
  \country{South Korea}
}
\email{marko.k@kakaocorp.com}

\renewcommand{\shortauthors}{Kim et al.}

\begin{abstract}
  We present an effective way to predict search query-item relationship. We combine pre-trained transformer and LSTM models, and increase model robustness using adversarial training, exponential moving average, multi-sampled dropout, and diversity based ensemble, to tackle an extremely difficult problem of predicting against queries not seen before. All of our strategies focus on increasing robustness of deep learning models and are applicable in any task where deep learning models are used. Applying our strategies, we achieved 10th place in KDD Cup 2022 Product Substitution Classification task.
\end{abstract}



\maketitle

\section{Introduction}
KDD Cup 2022 presents a problem of having to accurately describe
the relationship between user search queries and shopping items,
in motivation to improve customer experience and search engagement.
The competition presents Shopping Queries Dataset \cite{reddy2022shopping},
a multilingual e-commerce dataset in three languages:
English, Spanish and Japanese.

Among the three tasks presented in the competition,
the goal of “Product Substitute Identification” (task 3)
is to correctly identify whether a search query and and an item are
in a substitute relationship or not.
A relationship is considered substitute when an item fails to
fulfill some aspects of the query but can be functioned as substitute of a completely 
relevant item. The competition’s evaluation metric is Micro F1 Score as classes are
imbalanced.

In this work, we discuss a simple but effective model architecture and training techniques that placed us among top-10 solutions in KDD Cup 2022. Built upon a solid validation strategy, where models are evaluated based on “unseen” query - item pairs, our solution is a simple combination of pre-trained transformer \cite{NIPS2017_3f5ee243} \cite{DBLP:journals/corr/abs-1810-04805} and LSTM \cite {10.1162/neco.1997.9.8.1735}, trained using effective techniques to increase model robustness, such as exponential moving average of weights, multi-sampled dropout, dynamic learning rate scheduling for different layers. Our final solution is a mix of models fused with diversity based ensemble. As all our strategies are task-independent and focus on increasing robustness of deep learning models in general, we believe our methodologies are easily applicable to any other tasks where deep learning models are used.

\section{Proposed Solution}

\subsection{Redefining the Problem}

We realized that dataset had been split into train and public test set grouped by queries, so that queries in public test set do not appear in train set and vice versa. So we redefine the problem as identifying relationship between “unseen” queries and items, which makes the problem extremely difficult. Realizing this difference was crucial as wrong validation strategy may not be able to properly mimic the competition evaluation environment.

\begin{table*}
  \caption{Model Performance for each Technique Applied}
  \label{tab:performance}
  \begin{tabular}{ccl}
    \toprule
    Technique & Cross Validation Score & Relative Gain/Loss vs. Control\\
    \midrule
    DeBERTa-v3-Large (Baseline) & 0.8229 & +0.0000 \\
    Add LSTM Head & 0.8226 & -0.0003 \\
    Add LSTM Head with higher LR & 0.8234 & +0.0005 \\
    Adversarial Training & 0.8262 & +0.0028 \\
    Exponential Moving Average & 0.8265 & +0.0003 \\
    Multi-Sampled Dropout & 0.8267 & +0.0002 \\
    Cosine Schedule → StepLR & 0.8267 & +0.0005 \\
    XLM-RoBERTa-Large & 0.8237 & -0.0026 \\
    RemBERT & 0.8247 & -0.0030 \\
    \bottomrule
  \end{tabular}
\end{table*}

\subsection{Validation Strategy}

When dataset is split into a regular random train-test split, model training worked relatively well and validation loss would successfully decrease for 8 to 10 epochs. However, same model, trained the same way, on a train-test split where sets of queries (ex. airpods) would appear only on either train or test set, model training will not work well and validation loss would not decrease for a single epoch. We presume this occurs as the given problem is not a general text classification task but requires identification of two texts: search query and item. So we decided to split the given train dataset in the same manner.

As there exists class imbalance between substitute labels and non-substitute labels, we use stratification to ensure equal distribution of each class in local train and test sets. To sum up, we use StratifiedGroupKFold cross validation, where K is 5, and grouped on queries, and determined model performance solely upon micro f1 score on all folds.

All of our experiments are run in a strict comparison environment, where there always exists a control model for comparison and only a single hyper-parameter would be changed for the experiment model.

\subsection{Model Backbone}

We started out the competition, after some exploratory data analysis, with an agressive model selection. We tried out both classic LSTM models and pre-trained transformers and the latter scored around 0.05 higher in previously mentioned cross validation setting. Including the multilingual-BERT model \cite{DBLP:journals/corr/abs-1810-04805} used as competition baseline model, we tried out many models including XLM-RoBERTa \cite{DBLP:journals/corr/abs-1911-02116}, RemBERT \cite{DBLP:conf/iclr/ChungFTJR21} to check basic model performance. DeBERTa-V3-Large \cite{he2021debertav3} came out to be the most effective pre-trained model by far, achieving micro-f1 score of 0.8229, and selected as our baseline model.

For model input, we concat search query and item titie using special token in following manner: “<Query> [Special Token] <Item Title>”. Item title goes through a very simple preprocessing of only cleaning special characters. More intense cleaning, such as underscoring, deteriorated model performance. Maximum sequence length is set to 78, covering approximately 99.7 percent of all data samples.

\begin{figure*}[t]
\centering
\includegraphics[width = 1\textwidth]{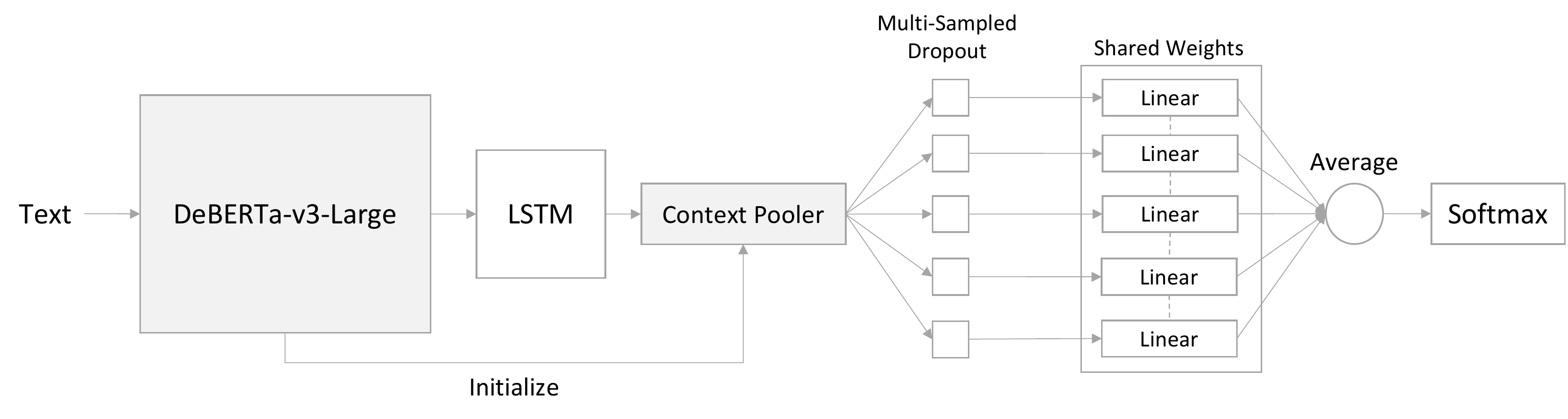}
\caption{Visualization of Final Model Structure}
\label{fig:finalmodel}
\end{figure*}

\subsection{LSTM Head}

We improve our model structure by adding an LSTM layer after DeBERTa backbone. Note that the added LSTM layer is not pre-trained. As training finished in one or two epochs, applying DeBERTa-V3-Large’s learning rate of 5e-6 was not adequate and actually led to a decrease of -0.0003 in f1 score. So the LSTM layer would need higher learning rate than other pre-trained layers. We initially set learning rate for pre-trained DeBERTa layers to 5e-6 and LSTM layer to 1e-3 and use learning rate scheduling to decrease the rates during model training. Applying LSTM head with adjusted learning rate led to a f1 score gain of +0.0005.

\subsection{Adversarial Training}

We apply Adversarial Weight Perturbation \cite{10.5555/3495724.3495973} to increase model robustness. We use adversarial learning rate of 1e-4. Starting adversarial training from the very beginning was most effective, instead of starting at a later period (ex. epoch 2), as entire training ended in a very short period. After applying Adversarial Weight Perturbation, training became much more stable: validation loss would decrease for 3 to 4 epochs instead of a single epoch. This led to a f1 score gain of +0.0028.

\subsection{Exponential Moving Average}

We calculated exponential moving average of weights during mini-batch training. We applied a decay of 0.999 when calculating new average of weights. Doing so led to a relative micro-f1 score gain of +0.0003.

\subsection{Multi-Sampled Dropout}

We apply Multi-Sampled Dropout \cite{https://doi.org/10.48550/arxiv.1905.09788} for better generalization. After DeBERTa’s pooling layer, we add 5 dropout layers with different masks, each with a dropout rate of 0.5. Output of each dropout layers will be passed to fully connected layers, which all share the same weights. Output of all fully connected layers were averaged.

At this point, the final model structure is shown in Figure 1. It is a simple model with a pre-trained DeBERTa backbone, LSTM head, pooling layer initialized using pre-trained DeBERTa weights, multi-sampled dropout layers, and finally the classification layer.

\subsection{Learning Rate Scheduling}

We used cosine learning rate scheduling throughout the competition. Towards the end of competition, we tried out StepLR schedule with decay every epoch, decayed by gamma of 0.2. Changing learning rate scheduler resulted a +0.0005 change in competition metric.

\subsection{Diversity based Model Ensemble}

After applying all modifications mentioned above, we re-trained the top few backbone models for a secondary model selection, to check whether our modifications would impact some backbone models to a greater degree and to use some of the models in ensemble. As a result, the RemBERT model, which had a -0.003 f1 score loss in the original model selection, came out close second with only 0.0003 score difference with DeBERTa-V3-Large. Although applying both 2.6 Exponential Moving Average and 2.7 Multi-Sampled Dropout led to a total of +0.0005 score gain for DeBERTa-V3-Large model, we can see that applying the same strategies on the RemBERT model led to much greater score gain.

In addition to a soft-voting based ensemble, a technique where model prediction probabilities are summed to get final predictions, we use a diversity based approach when ensembling different models. For model diversity, instead of using the same backbone twice during model ensemble, it was shown to be more effective to use two different backbones, even if it’s single model score was relatively lower. Along with DeBERTa-V3-Large and RemBERT models, using the XLM-RoBERTa-Large model, despite its lower f1 score, in the end led to both highest public and private leaderboard scores among all ensemble combinations.

For data diversity, we ensured each of our ensemble models were trained on different datasets. We realized two days before the end of competition that there was some non-overlapping data between task1 and task 3. As more data can make deep learning models more robust, we secured task 1’s non-overlapping 234,286 samples of data, which is a 12.8 percent increase in the numbers of total train data. Table 2 shows cross validation scores for the combined dataset. Note that cross validation scores decrease when evaluated on the combined dataset which does not mean decrease in model performance but a change due to the extra data. In the end, we selected DeBERTa-V3-Large validated on fold 0 and trained on the rest, and RemBERT model validated on fold 2 and trained on the remaining folds. Unfortunately, we had to use XLM-RoBERTa-Large model trained only on task 3 data and could not apply some techniques such as multi-sampled dropout and exponential moving average, due to time constraints.

\begin{table}[h]
  \caption{Final Submitted Models Summary}
  \label{tab:summary}
  \begin{tabular}{cccccccl}
    \toprule
    Model & CV (Task 1+3) & 2.4 & 2.5 & 2.6 & 2.7 & 2.8 \\
    \midrule
    DeBERTa-v3-Large & 0.8165 & O & O & O & O & O \\
    RemBERT & 0.8162 & O & O & O & O & O \\
    XLM-RoBERTa-Large & - & O & O & X & X & O \\
    \bottomrule
  \end{tabular}
\end{table}

Table 2 describes our set of final models. In summary, our final submission uses three backbone models: DeBERTa-v3-Large, RemBERT, and XLM-RoBERTa-Large. For the first two models, both task 3 and task 1 data were used for training.

\subsection{Accelerated Inference}

The competition submission environment used a single NVIDIA V100 GPU (16GB RAM) and had a timeout of 120 minutes for docker environment setup and both public and private test set evaluations. To maximize the number of models used for ensemble, we applied float16 precision, which improved inference speed by 2.3 times. More specifically, we were able to reduce inference time of public test set’s 277044 samples for a DeBERTa-v3-Large model, from 23 minutes to 10 minutes when sequence length was 78. For sequences of length 256, inference time decreased from 82 minutes to 36 minutes.

\begin{table}[h]
  \caption{Inference Times for Different Float Precisions}
  \label{tab:inference}
  \begin{tabular}{cccccccl}
    \toprule
    Sequence Length & Float32 Precision & Float16 Precision \\
    \midrule
    78 & 23 minutes & 10 minutes \\
    256 & 82 minutes & 36 minutes \\
    \bottomrule
  \end{tabular}
\end{table}

\section{Conclusion}

In this work, we discussed our approach of effectively capturing the relationship between search queries and shopping items, as part of our 10th place solution in KDD Cup 2022. Built upon a solid validation strategy, where models are evaluated based on “unseen” query - item pairs, our solution is a simple combination and pre-trained transformer backbone and LSTM head, trained using techniques to increase model robustness, including adversarial training, exponential moving average, multi-sampled dropout, dynamic learning rate scheduling for different layers. Our final solution is an ensemble of models selected based on both model and data diversity. We believe our methods presented in this paper would help increase robustness of deep learning models in general.

\bibliographystyle{ACM-Reference-Format}
\bibliography{sample-base}

\end{document}